\documentclass[journal=jacsat,manuscript=article]{achemso}
\usepackage[version=3]{mhchem}
\usepackage[font=small,labelfont=bf]{caption}
\usepackage{siunitx}
\usepackage{enumerate}
\usepackage{setspace}
\usepackage{pifont}
\usepackage[table]{xcolor}
\usepackage{color,soul}
\usepackage{amssymb, bm}

\usepackage{setspace}
\usepackage{algorithm, algpseudocode}
\usepackage{lineno}
\usepackage[backgroundcolor=blue!20!white,textsize=tiny,textwidth=2.1cm]{todonotes}
\usepackage{multirow}
\usepackage{tcolorbox}
\usepackage[dvipsnames]{xcolor}

\author{Hannes Kneiding,$^{\dagger}$ Lucía Morán-González,$^{\dagger}$ Nishamol Kuriakose,$^{\dagger}$ Ainara Nova,$^{\dagger}$ David Balcells,$^{\dagger,}$}

\email{david.balcells@kjemi.uio.no}

\affiliation{$^{\dagger}$Hylleraas Centre for Quantum Molecular Sciences, Department of Chemistry, University of Oslo, P.O. Box 1033, Blindern, 0315 Oslo, Norway}

\title{Inverse Design of Inorganic Compounds \\ with Generative AI}

\begin{document}

\thispagestyle{empty}

\maketitle

\newpage

\thispagestyle{empty}

\begin{center}
    \includegraphics[scale=3.00]{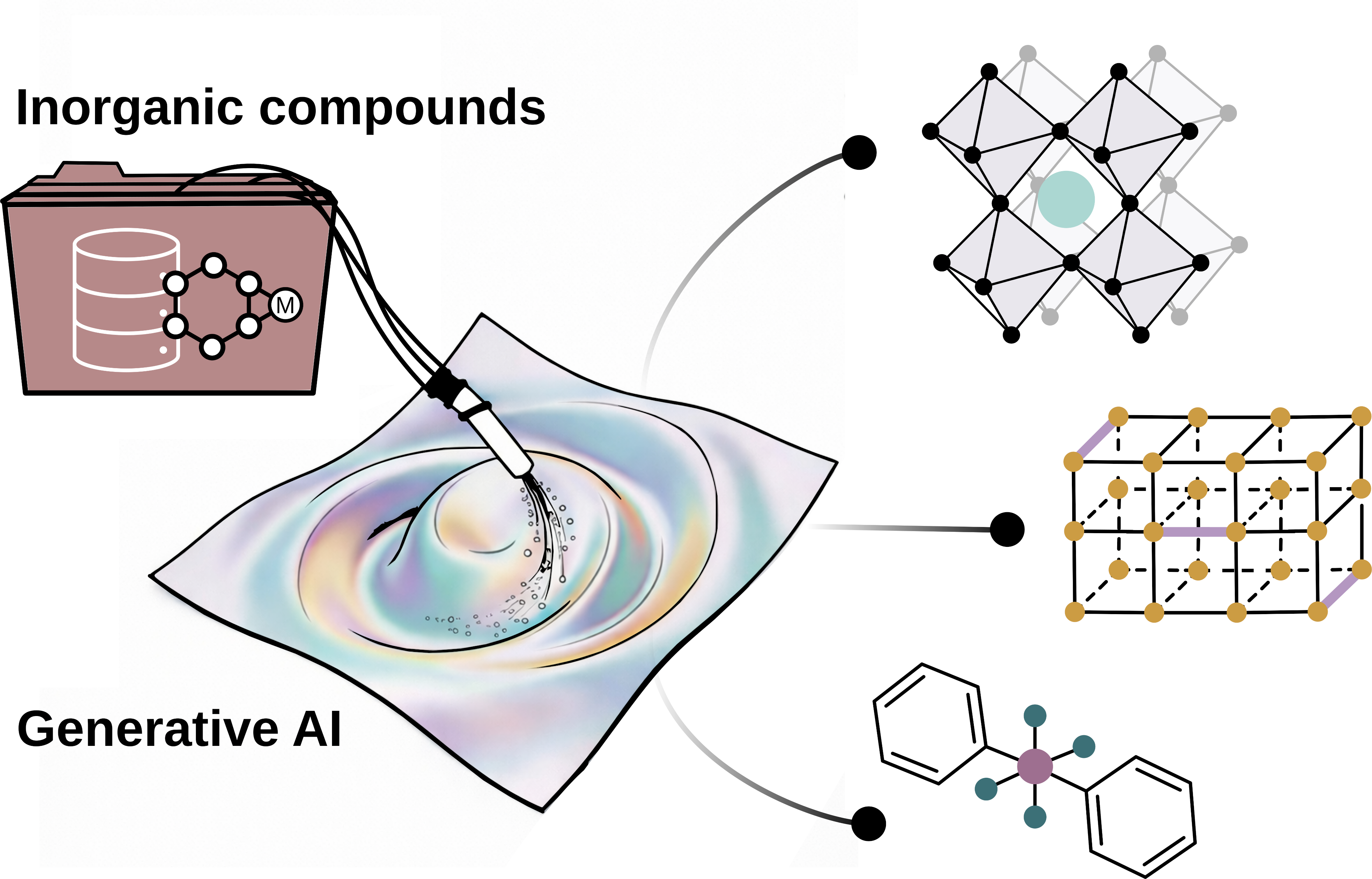}
\end{center}

\subsection{Abstract}

Machine learning is revolutionizing chemistry. Beyond the value of predictive models accelerating virtual screening, generative AI aims at enabling inverse design, reversing the compound-to-property prediction paradigm into property-to-compound generation. Chemists now have access to a rich AI toolbox for organic chemistry, including drug discovery. However, the application of these methods to inorganic compounds remains limited by the challenges posed by their intrinsic nature. This Perspective analyzes how these challenges have been addressed, considering widely diverse systems ranging from molecules to crystals, including transition metal complexes and microporous materials. The analysis focuses on how generative AI methods have evolved towards data-representation-model pipelines that address the full complexity of inorganic compounds, including their chemical composition, geometry, symmetry, and electronic structure. Future directions, like benchmark standardization and the development of synthesizability metrics, are also discussed.

\newpage

\section*{Introduction}

\bigskip

\noindent
The development and application of inorganic compounds has been driven by both empirical discovery and rational design. The past century witnessed a paradigm shift with the advent of coordination chemistry, followed by solid-state theory and the concept of tailored porosity. \\

\begin{figure}[H]
    \centering
    \includegraphics[scale=0.14]{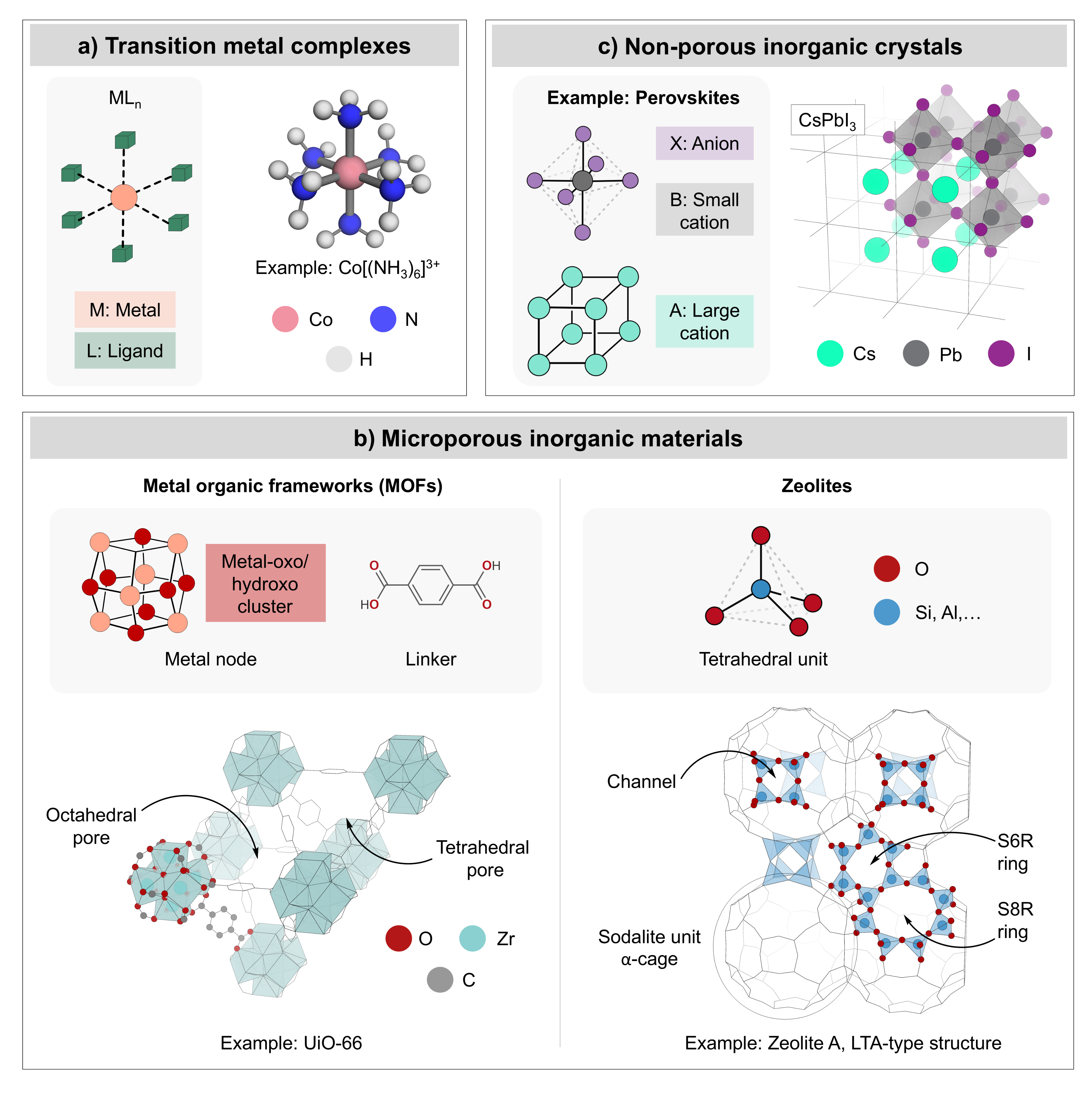}
    \caption{Inorganic compounds covered in this Perspective. \textbf{a)} Transition metal complexes are built around a metal center supported by a set of ligands. \textbf{b)} Microporous inorganic materials have porous reticular structures made with metal node-linker (MOFs) or aluminosilicate (zeolites) building blocks. In MOFs, the metal node can also be a single-atom cation. \textbf{c)} Non-porous inorganic crystals form a compact, periodic structure made of metals or ions.}
    \label{fig:Introduction}
\end{figure}

\noindent
The foundations of coordination chemistry were laid in 1893 with the seminal work of Alfred Werner, who introduced a structural description of metal–ligand bonding.\cite{Werner267} This concept laid the foundation of transition metal complexes (TMCs), in which a metal center is bound to a set of ligands (Figure \ref{fig:Introduction}a). The diversity of molecular architectures arising from the ligands and the metal coordination number, geometry, and oxidation state has been leveraged across a remarkably broad spectrum of applications. Modern organic synthesis relies on TMC‑catalyzed asymmetric hydrogenation\cite{Knowles1445,Miyashita7932} and C–C cross‑coupling reactions,\cite{Miyaura3437,Negishi1821,Heck2320,Docherty7692} which are deployed routinely from micromole-scale discovery campaigns to multi-kilogram manufacturing processes.\cite{Ganley8317} Beyond this central role, TMCs have yielded clinically-tested metallodrugs,\cite{Rosenberg5191} mediated C–H bond activation,\cite{Milan9559} and driven advances in renewable energy technologies.\cite{McWilliams2020,Liange202200723} Major improvements in characterization techniques and computational chemistry have enabled electronic modeling of metal–ligand interactions.\cite{Goswami719, Maurer6134} This deeper knowledge has, in turn, driven developments in photoredox catalysis,\cite{Prier5322, Zhiwei437, Chan1485} in the design of bimetallic systems,\cite{Navarro11220} and in the comprehensive elucidation of complex TMC-catalyzed pathways.\cite{Vogiatzis2453} Current frontiers of TMC chemistry are catalysis with earth-abundant metals,\cite{Wheelhouse1157} next-generation metallodrugs,\cite{Kim100569} and applications to quantum information science.\cite{Sauza12896} \\

\noindent
Early crystallographic techniques revealed atomic arrangement of solids, but its rational engineering initially remained elusive. This began to change with the development of reticular chemistry, in which building blocks are assembled into extended frameworks with controlled porosity.\cite{Jaheon705} Metal organic frameworks (MOFs) have since emerged as a prototypical class of modular crystalline microporous solids (Figure \ref{fig:Introduction}b). Their architectures allow systematic variation of the pore size, geometry, and chemical functionality without altering the underlying network topology, a conceptual breakthrough recognized by the 2025 Nobel Prize in Chemistry.\cite{Hoskins1546, Kondo1725, Eddaoudi469} The combination of exceptional tunability and high internal surface area has made MOFs prominent candidates for gas capture\cite{Zhou96, Lin1464, Daglar2405532, Fathieheaat3198, Furukawa1230444} and storage,\cite{Alezi13308, Farha693, Chen297} as well as for liquid-phase molecular separations,\cite{Sini92} all of which are widely pursued in energy-related technologies.\cite{Asgari4701} Although early examples often lacked durability, chromium and zirconium-based frameworks\cite{Ferey2040, Cavka13859} combining large-volume pores with high stability enabled the use of MOFs under harsh electrochemical\cite{Xu292} and catalytic\cite{Rajenahally326, Yang1779} conditions. \\

\noindent
Long before reticular chemistry was formalized, related ideas about controlled porosity were explored in the realm of zeolites.\cite{Barrer2158} Unlike MOFs, zeolites are built from a more restricted palette of aluminosilicate tetrahedral frameworks (Figure \ref{fig:Introduction}b). Individual structures exhibit fixed channel dimensions and are classified by a three-letter code that denotes the underlying topology. Rigid SiO\textsubscript{4}/AlO\textsubscript{4} networks, combined with tunable Brønsted and Lewis acidity, confers robustness and pronounced shape selectivity,\cite{Corma559,Chizallet6107} positioning zeolites as a workhorse for large-scale separation processes and heterogeneous catalysis in the petrochemical industry.\cite{Li1156, PerezBotella17647, Campo8511, Jiao727, Matito19884} Contemporary synthetic strategies emphasize greener routes,\cite{Deng29115, Abdel175} compositional tuning though isolated metal single sites\cite{Zhang6039,Zhao7174} and the creation of intrinsic meso/micro-porous channel systems.\cite{Lu368} In parallel, computational methodologies are integrating solid-state and molecular descriptors of catalysis\cite{Speybroeck7044} in the design of zeolites as nanoreactors, considering  diffusion and confinement effects.\cite{Yan7103} \\

\newpage

\noindent 
Whereas MOFs and zeolites derive much of their utility from internal surface area, non-porous inorganic crystals achieve their functionality through dense, long-range periodic packing. Dense inorganic materials such as alloys\cite{JienWei471} and 2D layer materials\cite{Vladan15891} combine robustness with high charge-carrier mobilities arising from continuous orbital overlap. Perovskites, with the general formula $ABX_3$ (Figure \ref{fig:Introduction}c), simultaneously offer defect tolerance, band gap tunability, and various optical properties.\cite{Kojima6050, Sinkeaad4424, Snaith372} These attributes have fostered their rapid adoption in photovoltaic devices,\cite{Im4088, Chiba681} X-ray detectors,\cite{Song1056} and electrocatalysts.\cite{Hwang751} Current endeavors aim to push single-cell efficiencies towards the single-junction theoretical limit.\cite{Amran1300, Lee643, Leeeabj1186} Complementary efforts target improvements in long-term operational stability\cite{Snaith372, CorreaBaena739, Zhang194} and mitigating the risks associated with lead leakage into the environment.\cite{Chen2204144} \\

\noindent
The further development of inorganic compounds is urged by the crises caused by increasing energy demands, antimicrobial resistance, and climate change. At the core of the technologies addressing these challenges, there is a need for more active and selective catalysts\cite{Moran9089, Xin102} and metallodrugs,\cite{Wang12361} as well as more efficient microporous and crystalline materials for energy storage and transformation.\cite{Eliyahueadn9391, Yu30568} In this context, machine learning (ML), and more generally, artificial intelligence (AI), are accelerating the study and discovery of novel inorganic compounds. In contrast with \texttt{discriminative AI}, which focuses on compound-to-property predictions, \texttt{generative AI} enables the \texttt{inverse design} of novel molecules and materials via the reverse property-to-compound approach.\cite{Anstine8736, Noh4871} \\

\noindent
This Perspective analyses the application of generative AI to inorganic compounds, with sections dedicated to TMCs, non-porous inorganic crystals, and microporous inorganic crystals. Aiming at the interdisciplinary community that works on the development and application of generative AI, the closing outlook discusses the capabilities and limitations of these emerging methods in inverse design, defining application boundaries and synergy opportunities. The next section provides an overview of the most relevant methods. \\

\section*{Generative AI machinery}

\bigskip

\noindent
Generative AI methods for inverse design (Box 1) can be classified into \texttt{deep learning} (DL) or \texttt{evolutionary computing} (EC). DL methods use \texttt{neural networks} to learn an internal representation, the \texttt{latent space}, from which compounds are generated. State-of-the-art methods include diffusion models (DMs),\cite{Chen2024} variational autoencoders (VAEs),\cite{Kingma2019} and large language models (LLMs).\cite{Xin2026} DL methods like generative adversarial networks (GANs)\cite{Gui3313} and deep reinforcement learning,\cite{Murphy2025, Zhou10752} have seen less development in the field, though RL, for example, can be a powerful tool when used within other generative models like DMs.\cite{Chen2025} Generative AI methods allow for the exploration of the chemical space\cite{Restrepo568} via the random generation of compounds (\texttt{unconditional generation}) beyond the million-system scale, followed by \texttt{high-throughput virtual screening (HTVS)} to validate their properties. However, given the nearly infinite size of the inorganic compound space, inverse design by \texttt{conditional generation} is preferred, since it directly steers generation towards the fewer compounds that meet the target properties of interest. These properties can be, for example, energy barriers and band gaps in the inverse design of catalysts and energy materials, respectively. \\

\begin{tcolorbox}[width=\textwidth, colback=white, title={\textbf{Box 1. Selected generative AI methods: Deep learning}}, colbacktitle=Maroon, colframe=Maroon, coltitle=white, arc=2mm, boxrule=0.75mm, boxsep=1.75mm]

\bigskip

\noindent
\textbf{Diffusion model (DM).} A model that, through a controlled stochastic process, transforms a compound representation into noise, then using neural networks to learn the reverse denoising process that recovers the representation. The noising and denoising processes are sequential and stepwise. For inverse design, the denoising steps take into consideration both the representation and the property of interest in a conditional probability distribution, which, after training the model, allows for guiding the process towards regions of chemical space that satisfy target property constraints. Flow models can be seen as a DM variation in which the denoising is done with learned differential equations, instead of stochastic processes. \\

\noindent
\textbf{Generative adversarial network (GAN).} A model based on two competing neural networks: one producing compounds from random noise (\emph{generator}), and one distinguishing these compounds from the training data distribution (\emph{discriminator}). GANs are trained iteratively towards a state in which the distribution of generated compounds becomes indistinguishable from that of the training data (\emph{Nash equilibrium}). For inverse design tasks, the model is also fed with known values of the target property, allowing to condition the generator to this property.

\end{tcolorbox}

\begin{tcolorbox}[width=\textwidth, colback=white, title={\textbf{Box 1 (continued). Selected generative AI methods: Deep learning}}, colbacktitle=Maroon, colframe=Maroon, coltitle=white, arc=2mm, boxrule=0.75mm, boxsep=1.75mm]

\bigskip

\noindent
\textbf{Large language model (LLM).} A model that uses a complex neural network (\emph{transformer}) to process chemical information in string formats like SMILES. With a mechanism that is aware of the whole string as it is generated (\emph{self-attention}), LLMs predict the next fragment (\emph{token}) of a string that defines a chemical compound (\emph{auto-regression}). General purpose LLMs can be re-trained (\emph{fine-tuned}) with datasets providing specific information; for example, properties of interest for a large set of inorganic compounds. For inverse design, LLMs are guided with engineered prompts (\emph{chain-of-thought}) building a conversation (\emph{context window}) in which a chemist requests the generation of compounds with desired properties. In agentic AI systems, LLMs are also used as `wrappers' holding other generative AI models at their core, for example, GAs, acting as a communication interface to the user. \\

\noindent
\textbf{Variational autoencoder (VAE).} A model that first learns the encoding of an input compound into an internal representation (\emph{latent space}) and then learns its decoding back to the reconstructed, original compound. Both the encoder and decoder are often graph neural networks that are trained to maximize the quality of the reconstruction, while regularizing the latent space towards a normal distribution. For inverse design, an additional neural network is added and trained jointly with the encoder and decoder to map the latent space onto a property of interest. This allows for navigating the latent space to find and decode compounds that have the targeted property values.

\end{tcolorbox}

\begin{tcolorbox}[width=\textwidth, colback=white, title={\textbf{Box 1 (continued). Selected generative AI methods: Evolutionary computing.}}, colbacktitle=Maroon, colframe=Maroon, coltitle=white, arc=2mm, boxrule=0.75mm, boxsep=1.75mm]

\bigskip

\noindent
\textbf{Genetic algorithm (GA).} An optimization algorithm that, through multiple iterations (\emph{populations}), replicates and diversifies the modular representation of a chemical compound (\emph{genes} in \emph{chromosome}) using random change (\emph{mutation}) and exchange (\emph{crossover}) operations. No training data is required, but rather generated on-the-fly, as the chemical space is explored generation after generation. Evolution is driven by the selection of the fittest compounds in each generation (\emph{survivors}), according to a fitness function of the target properties. A probabilistic selection of survivors define the compounds (\emph{parents}) that yield the next generation (\emph{offspring}) through the genetic operations. Inverse design is intrinsic to the gradient-free optimization of the fitness towards extreme values. Regions of joint optimality for multiple targets (Pareto front) can also be explored via multi-objective optimization.

\end{tcolorbox}

\bigskip

\noindent
In EC methods,\cite{Dan2013} inverse design is intrinsic to the optimization of a \texttt{fitness function} that evaluates how well candidate solutions satisfy the target properties. Genetic algorithms (GAs), which are one of the most established EC methods (Box 1), perform this task via genetic operations together with a selection process that pushes a population of candidates towards maximal fitness function values. GAs are particularly suited for inorganic compounds due to their modular nature, allowing for a straightforward definition of genetic operations that change and exchange the building blocks; for example, in MOFs, the metal nodes and organic linkers (Figure \ref{fig:Introduction}b). In computer science, GAs are classified as optimization algorithms, but in the context of chemistry and materials science, they are also often considered as generative methods\cite{Anstine8736, Xie2336} since both approaches serve essentially the same purpose: to deliver chemical systems that meet one or more target properties, which, in the case of GAs, achieve optimal (minimum or maximum) values, as it is normally desired. \\

\noindent
Data and representations are also critical for the development of the AI models summarized in Box 1. The quality of the data, including its correctness and scope, is essential, since it sets the quality limit of the model in which it is used for training. While EC methods like GAs do not require training data, they still require data to define the building blocks manipulated in the genetic operations. Though still scarcer than the data available for organic compounds, there is a growing library of inorganic compound datasets that can be used for generative AI, covering extensive and diverse regions of the TMC, MOF, zeolite, and inorganic crystal spaces (Table \ref{tab:datasets}). Many datasets are computational, requiring methods that account for specific properties found in inorganic compounds like, for example, the relativistic effects introduced by heavy elements. Datasets like tmQM\cite{Balcells6135, Kneiding618, Kneiding11766} can be used in full, whereas others, like the MP,\cite{Jain011002} are used to extract subsets for training tailored models. Further, datasets like the OMat24\cite{Barroso2024} were originally developed for training machine learning interatomic potentials (MLIPs), but they contain both non-equilibrium and equilibrium structures that can be used to train generative AI models.\\

\noindent
Generative AI requires machine-readable representations of chemical compounds both at the input level, to train the models, and at the output level, for their generation. These representations express composition and structural information, which, in line with the nature of the datasets (Table \ref{tab:datasets}), can be either experimental, like X-ray diffracted structures (CIF files; CIF = crystallographic information file), or computational, like geometries optimized via density functional theory (DFT) calculations (\texttt{XYZ files}, with the atomic $x,y,z$ Cartesian coordinates). Relative to organic molecules, inorganic compounds challenge their own representation with higher complexity in chemical composition, electronic structure, and geometry, including a wider set of elements, $d$ and $f$ orbital contributions to bonding, and periodicity subject to symmetry constraints. Representations commonly used for organic chemistry, like SMILES and graphs, are more difficult to formulate, and the associated chemoinformatics tools, for example RDKit, have limited applicability. Further, representations for generative AI need to be invertible, since they are both the model input and output. Table \ref{tab:reps} provides a descriptive selection of inorganic compound representations for generative AI. The inclusion of chemical data in them, and its physics-informed use within the models, boost their performance in inverse design tasks. \\

\begin{table}[ht]
  \begin{center}
  \begin{footnotesize}
   \caption{Datasets available for training generative AI models for inorganic compounds.}
   \label{tab:datasets}
    \begin{tabular}{l l l l}
    \hline
    \textbf{Name}$^a$ & \textbf{Size} & \textbf{Contents} & Ref. \\
    \hline
    \textbf{CSD} & 1.4M & Experimental composition and structure data & \cite{Groom171} \\
    \textbf{COD} & 0.5M & Experimental composition and structure data & \cite{GražulisD420} \\
    \textbf{ICSD} & 328k & Experimental \& computational composition and structure data & \cite{Zagorac918} \\
    \textbf{PCD} & 395k & Experimental \& computational composition and structure data & \cite{Villars2024} \\
    \textbf{MP} & 785k & Computational structure, quantum properties, and thermodynamic data & \cite{Jain011002} \\
    \textbf{OQMD} & 1.4M & Computational structure, quantum properties, and thermodynamic data & \cite{Saal1501} \\
    \textbf{OC22} & 62k & Computational data for metal-based heterogenous catalysts & \cite{Tran3066} \\
    \textbf{OMat24} & 110M & Computational structure, quantum properties, and thermodynamic data & \cite{Barroso2024} \\
    \textbf{OMol25} & 83M &  TMC computational structure, quantum, and thermodynamic data & \cite{Levine2025} \\
    \textbf{tmQM} & 108k &  TMC computational structure and quantum properties data & \cite{Balcells6135, Kneiding618, Kneiding11766} \\
    \textbf{CoRE} & 40k & MOF experimental \& computational, composition, and structure data & \cite{Chung5985, Zhao102140} \\
    \textbf{hMOF} & 138k & MOF computational structure data & \cite{Wilmer83} \\
    \textbf{L2M3} & 40k & MOF experimental synthesis and properties data & \cite{Kang3943} \\
    \textbf{Zeo-1} & 33k & Zeolite computational structure and quantum properties data & \cite{Komissarov61} \\

    \hline
    & & & \\
    \end{tabular}

    \noindent
    $^a$Acronyms: CSD = Cambridge Structural Database, COD = Crystallographic Open Database; ICSD = Inorganic Crystal Structure Database, PCD = Pearson's Crystal Data, MP = Materials Project, OQMD = Open Quantum Materials Database, OC22 = Open Catalyst 2022, OMat24 = Open Materials 2024, OMol25 = Open Molecules 2025, tmQM = transition metal Quantum Mechanics, CoRE = Computation-Ready Experimental, hMOF = hypothetical MOF, Zeo = Zeolite.
    \end{footnotesize}
  \end{center}
\end{table}

\noindent
While predictive ML models can simply be evaluated by measuring prediction errors relative to the ground truth values of the training data, the quality of generative AI models is more difficult to assess. In principle, a generated inorganic compound is only valuable if it is unique, novel, and valid. The stability of the generated compounds, their diversity, and the ease of making them in the experimental lab, are also important factors, among others. Due to the challenges in their representation, AI generated inorganic compounds are more difficult to evaluate compared to organic molecules. In the Outlook of this Perspective, we propose a collection of metrics (Box 2) that can be used to construct systematic evaluation workflows for the development of standardized benchmarks. \\

\begin{figure}[H]
    \centering
    \includegraphics[scale=0.40]{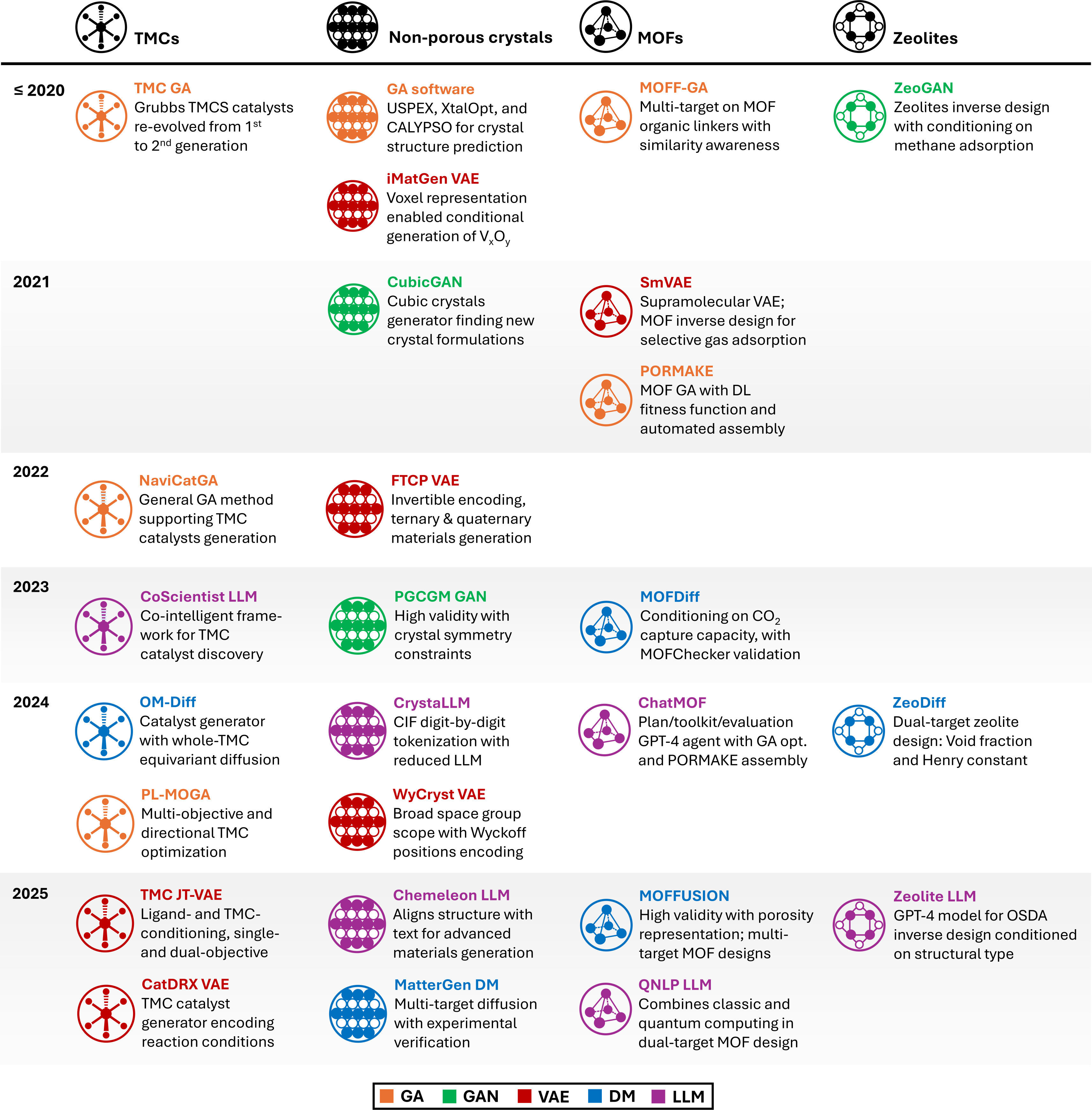}
    \caption{Timeline of selected generative AI methods, including genetic algorithms (GAs), generative adversarial networks (GANs), variational autoencoders (VAEs), diffusion models (DMs), and large language models (LLMs), for inorganic compounds, including transition metal complexes (TMCs), non-porous crystals, metal organic frameworks (MOFs), and zeolites.}
    \label{fig:timeline}
\end{figure}

\noindent
The evaluation of generative AI models for inorganic compounds has so far shown that DL methods require large training data ($\sim$10$^{3}$ and bigger) and complex representations ($\sim$10$^{2}$ dimensions) to achieve high chemical validity. Further, inverse design tasks conditioned by multiple target properties remains challenging. State-of-the-art DL supports advanced features, like the \texttt{Euclidean invariances} of the target properties, being also able to generate diverse, unique, and novel compounds at scale. EC methods have lower performance in this regard but, relative to DL, they reduce the need for data while operating with simpler representations and offering \texttt{multi-objective optimization} algorithms for advanced inverse design tasks conditioned on multiple targets. The advantages and disadvantages of DL and EC methods are thus complementary to a significant extent, providing synergy opportunities.\cite{Kneiding15522} The next sections discuss how the application of generative AI has emerged and evolved over different types of inorganic compounds (Figure \ref{fig:timeline}). \\

\begin{table}[ht]
  \begin{center}
  \begin{footnotesize}
   \caption{Selected representations for generative AI methods for inorganic compounds.}
   \label{tab:reps}
    \begin{tabular}{l l l l l l l}
    \hline
    \textbf{Representations} & \textbf{Format}$^a$ & \textbf{Connect.} & \textbf{Isomerism} & \textbf{3D geom.} & \textbf{Periodic} & \textbf{DL/GA} \\
    \hline
    \textbf{SMILES}$^b$ & String & Yes & Partial$^c$ & No & No & Both \\
    \textbf{SELFIES}$^b$ & String & Yes & Partial$^c$ & No & No & Both \\
    \textbf{SLICES} & String & Yes & Yes & Yes & Yes & Both \\
    \textbf{Graph}$^d$ & Graph & Yes & Optional$^e$ & Optional$^e$ & Yes  & Both \\
    \textbf{3D grid}$^f$ & Point-wise grid & No & Yes & Yes & Yes & DL \\
    \textbf{Voxel}$^f$ & Cubic grid & No & Yes & Yes & Yes & DL \\
    \textbf{Point cloud}$^f$ & Coordinates$^g$ & No & Yes & Yes & Yes & DL \\
    \textbf{Wyckoff}$^f$ & Symmetry points & No & Yes & Yes & Yes & DL \\
    \textbf{Chromosome} & Genes & Yes & Partial$^c$ & No & No & GA \\
    \hline
    & & & & \\
    \end{tabular}
    
    \noindent
    $^a$Mathematically: Vector (strings and genes), matrix (coordinates and symmetry points), or tensor (graphs and 3D grids); $^b$Limited support for inorganic compounds; $^c$Rotational conformers excluded; $^d$Difficult to build from 3D geometries; $^e$Can be included through node and edge attribution; $^f$Can leverage Euclidean invariances; $^g$Can be either Cartesian (molecules) or fractional (crystals).
    \end{footnotesize}
  \end{center}
\end{table}

\newpage

\newpage

\hrule

\begin{center}
    \textbf{Glossary}
\end{center}

\hrule

\bigskip

\noindent
\textbf{$\bullet$ Conditional generation.} A generative task in which the output of a deep learning model is biased to meet target values of one or multiple properties of interest.

\noindent
\textbf{$\bullet$ Convex hull.} Composition versus formation energy/atom diagram. The energy above the convex hull is often used to define a limit between the stability and meta-stability of inorganic materials.

\noindent
\textbf{$\bullet$ Crystal structure prediction (CSP).} Inverse design task that generates crystal structures minimizing the energy under optional composition and symmetry constraints.

\noindent
\textbf{$\bullet$ Deep learning (DL).} AI methodology that uses multi-layer neural networks to learn $X \rightarrow y$ (prediction) or $X \leftarrow y$ (inverse design) mappings. $X$ is a machine-readable representation of a given chemical compound that has a property of interest $y$.

\noindent
\textbf{$\bullet$ Discriminative AI.} $X \rightarrow y$ classification and regression ML methods in which chemical properties ($y$) are predicted from composition and structure representations ($X$).

\noindent
\textbf{$\bullet$ Euclidean invariances.} When the E(3) group (rotation, translation, and reflection) is applied to a molecule or material, its physical properties, including those targeted for inverse design, can either change (variant), remain (invariant), or change accordingly (equivariant).

\noindent
\textbf{$\bullet$ Evolutionary computing (EC).} Generative AI methodology that drives inverse design via optimization. It uses algorithms inspired by biological evolution, for example, genetic algorithms, to derive chemical compounds with optimal property values.

\noindent
\textbf{$\bullet$ Fitness function.} In genetic algorithms, the function that is optimized (normally maximized) to drive the evolution process. It can involve one or multiple target properties, depending on whether it is used for single- or multi-objective optimization.

\noindent
\textbf{$\bullet$ Generative AI.} $X \leftarrow y$ ML methods that generate compounds ($X$) either unconditionally, without explicit biases, or conditionally, with a bias towards target properties ($y$).

\noindent
\textbf{$\bullet$ Grand canonical Monte Carlo (GCMC).} A random sampling technique that explores the configurational space of a chemical system under constant temperature, volume, and chemical potential.

\noindent
\textbf{$\bullet$ Graph neural network (GNN).} A neural network that leverages chemical and physical information through an architecture that reflects the topology of molecular or crystal graphs. This information is learned via message-passing, in which the graph nodes (atoms) exchange information through the edges (bonds). Modern models can efficiently leverage Euclidean invariances.

\noindent
\textbf{$\bullet$ High-throughput virtual screening.} Screening technique in which a large, explicit chemical space is explored by filtering its compounds through thresholds or ranges applied to computed properties.

\noindent
\textbf{$\bullet$ Inverse design.} Design approach in which chemical compounds are derived from a property objective.

\noindent
\textbf{$\bullet$ Latent space.} The internal representation learned by deep learning models.

\noindent
\textbf{$\bullet$ Loss function.} A differentiable function that quantifies the deviation between what the model generates and the ground-truth known from the training data. AI models are \emph{trained} (that is, optimized) by minimizing this function.

\noindent
\textbf{$\bullet$ Multi-objective optimization.} Inverse design task solved via the optimization of multiple target properties towards maximum or minimum values.

\noindent
\textbf{$\bullet$ Neural network.} A mathematical model that leverages probability and optimization theories in a sequential stack of alternating linear and non-linear operations. Neural networks predict compound properties or representations using parameters, like weights and biases, that are \emph{learned} (that is, optimized) to minimize a loss function.

\noindent
\textbf{$\bullet$ Pareto front.} In multi-objective optimization, the set of generated compounds that have at least one property that is optimal (minimal or maximal) relative to all other compounds.

\noindent
\textbf{$\bullet$ SUN metric.} The stability, uniqueness, and novelty (SUN) metric provides a partial assessment of the quality of AI models in the generation of molecules and materials.

\noindent
\textbf{$\bullet$ Unconditional generation.} A generative task in which the output of a deep learning model is produced without any bias, within a learned distribution. \\

\hrule

\newpage

\section*{Transition metal complexes}

\bigskip

While generative AI in chemistry has largely focused on small organic molecules\cite{Anstine8736, khater2025generative} and drug discovery,\cite{Vamathevan463, Zhavoronkov2026} interest in molecular inorganic compounds, particularly TMCs,\cite{Nandy13973, Nandy9927} has grown due to their roles in catalysis, medicine, and renewable energy. There are significant differences between organic compounds and TMCs that need to be considered; for example, datasets that are fundamentally different in size and diversity, and different sources of error in the computational methods used to generate the training data.\cite{Nandy9927} The higher complexity of the TMC chemistry requires adapting methods developed for small organic molecules, a challenge that has shaped the field. This adaption must address multiple oxidation states, involving different charges and spin multiplicities, high valency, involving different coordination numbers and geometries, and their interplay. While GAs have the longest tradition in TMC inverse design, recent years have also seen increasing success with DL approaches, including VAE, LLM, and DM models (Box 1, Figure \ref{fig:timeline}). \\

\begin{figure}[H]
    \centering
    \includegraphics[scale=1.55]{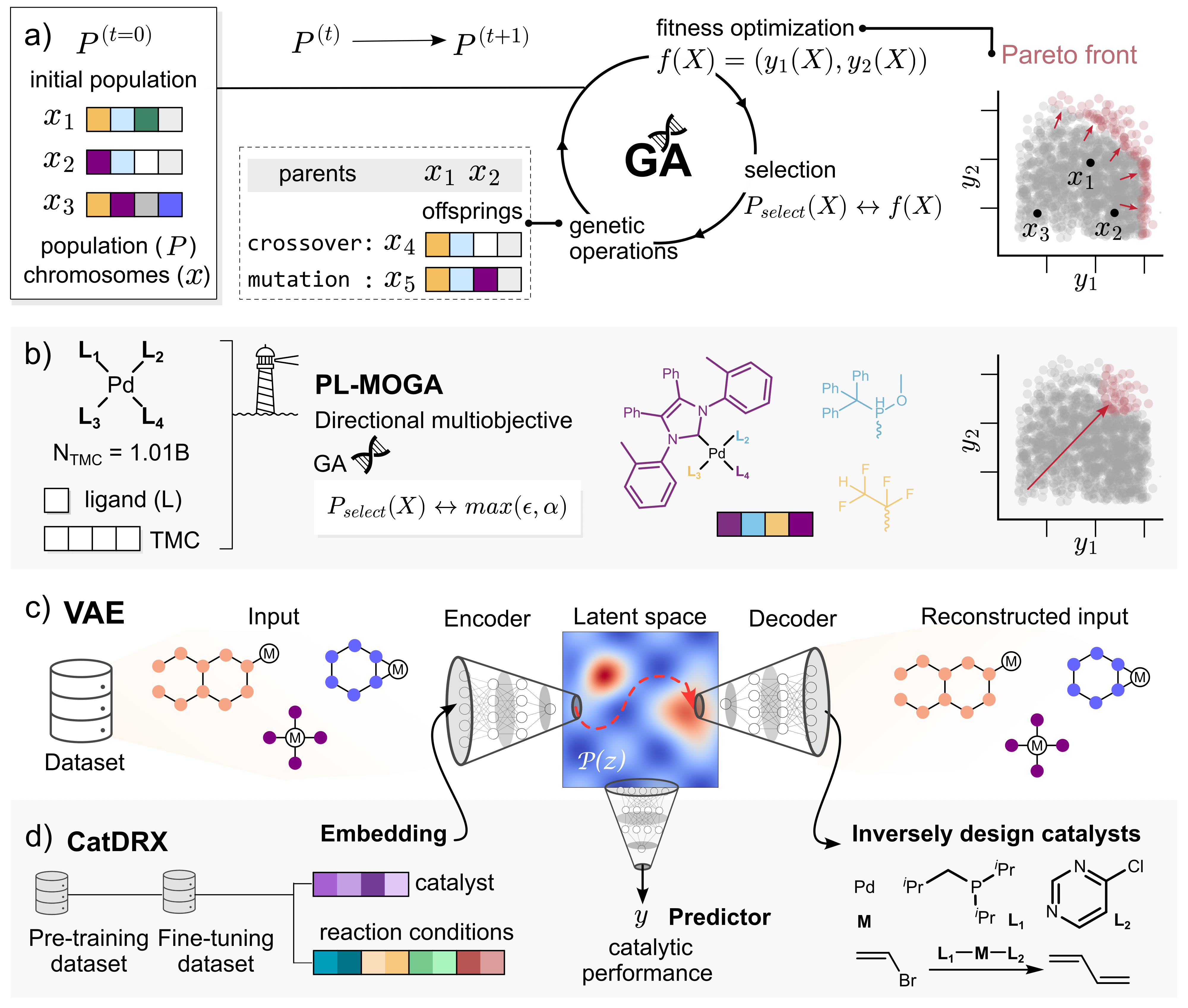}
    \caption{Selected generative AI methods for TMCs. \textbf{a)} GA using genetic operations, fitness optimization, and selection in Pareto front problems; \textbf{b)} The PL-MOGA GA for directional multi-objective optimization over selected regions of the Pareto front; \textbf{c)} VAE encoding and decoding metal ligands through a compressed latent space representation; \textbf{d)} The CatDRX VAE encoding both TMCs and reaction conditions for the inverse design of novel catalysts.}
    \label{fig:TMCs}
\end{figure}

\noindent
The metal and ligand blocks that form any TMC (Figure \ref{fig:Introduction}a) define a modular framework that is ideal for the application of GAs (Figure \ref{fig:TMCs}a). For inverse design, the coordination number and geometry of the metal center are fixed, while the ligand chemical space is explored by either combining  molecular fragments or by varying multiple, whole ligands. The latter approach is implemented via pre-defined ligand libraries like tmQMg-L.\cite{Kneiding263} This library has been used to explore massive chemical spaces; for example, with 252 ligands (merely 0.7\% of tmQMg-L), the combinatorics of the [PdL$_4$] scaffold yields 1.01 billion unique TMCs. GAs can explore these spaces efficiently while being easy to interpret through the analysis of the evolution process. However, by relying on a fixed library, they limit the novelty of the generated TMCs. Nonetheless, they constitute one of the most popular approaches in the literature, with reported applications including the optimization of metallocene catalysts for co-polymerization reactions,\cite{Kim202500316} spin-crossover\cite{Janet1064} and metal-oxo TMCs\cite{Nandy8243} for catalytic oxidation processes, and broad-spectrum, water-soluble chromophores for sustainable light-harvesting.\cite{pita2025evolving} Besides these specific applications, GAs that can operate across all possible metal and ligand coordination modes remain to be developed. \\

\noindent
In fragment-based GAs, ligands are built dynamically from predefined molecular fragments. The structure and diversity of these fragments determine how the ligands evolve during optimization in a process that extends novelty but requires moderation rules like, for example, setting a size limit. In the pioneering work of Chu and coworkers,\cite{Chu8885} TMCs were modeled as graphs divided into static and dynamic parts. The static part represents the metal center and its immediate surroundings, while the dynamic part contains substitution sites occupied by the predefined fragments. Since these fragments may contain further substitution sites, structure evolution can branch to explore diverse structures. Moderated mutation and crossover operations were able to re-discover second-generation Grubbs catalysts for olefin metathesis via the evolution of the first-generation, replacing the original phosphine ligands by carbenes. More recently, the NaviCatGA package generalized this approach, with proven success in other applications like the evolution of nickel TMC catalysts for the C--O bond cleavage of aryl ethers.\cite{Laplazae202100107} SMILES strings\cite{weininger1988smiles} can be used to simplify the implementation of the moderation constraints, treating each ligand as a separate molecule and defining the metal anchors on the basis of binding energies. \cite{strandgaard2023genetic, Strandgaard10638, seumer2025beyond} Offspring generation can then be implemented in terms of Jensen’s graph formalism,\cite{jensen2019graph} in which genetics are conveniently implemented as graph editing and subgraph exchange operations. This approach proved successful in the inverse design of catalysts for dinitrogen fixation\cite{strandgaard2023genetic, Strandgaard10638} and Miyaura-Suzuki cross-coupling reactions.\cite{seumer2025beyond} More robust string representations like SELFIES\cite{krenn2020selfies} have also proven their value in this context, using the first character to encode metal-binding, and substring-level genetics to evolve TMC molecular magnets.\cite{Frangoulis3808} \\

\noindent
To moderate the tendency towards complicated ligands, the synthetic accessibility score\cite{ertl2009estimation, coley2018scscore} can be added as an additional target, leveraging the multi-objective optimization abilities of GAs.\cite{strandgaard2023genetic, Strandgaard10638, seumer2025beyond} This can be implemented via an aggregated fitness function consisting in a product of exponential terms, each depending on a different target, which can be, for example, the stability and HOMO energy of metal-oxo TMCs.\cite{Nandy8243} This approach can also be leveraged to constrain evolution to specific regions of the chemical space,\cite{Janet1064,Nandy13973} but it has a relevant caveat: the need for defining sensible normalization factors to balance the optimization of the different targets. \texttt{Pareto front} optimization avoids this problem by treating the fitness as a set of multiple, independent targets. Only the non-dominated TMCs, which are those that are not outperformed by others across all targets, are kept, thereby enforcing a balanced optimization. The PL-MOGA algorithm implements this approach for TMCs,\cite{Kneiding263} adding the additional functionality of steering multi-objective optimization towards specific, user-defined regions of the Pareto front (Figure \ref{fig:TMCs}b). Besides the formulation of the fitness, GAs require the automated assembly of the TMC geometries, with tools like molSimplify\cite{Ioannidis2016molSimplify} or Architector,\cite{taylor2023architector} followed by their optimization and calculation of properties, often requiring quantum chemistry methods like DFT. Further, conformational search should be considered for flexible ligands and fluxional metal centers.\cite{strandgaard2023genetic, Strandgaard10638, seumer2025beyond} By representing TMCs with tailored features like graph autocorrelations,\cite{Janet8939, Moran8756} DFT ML-surrogates can be used to evaluate the fitness at a reduced computational cost.\cite{Janet1064, Nandy13973, Nandy8243, Frangoulis3808} In a broader context, the synergies between EC and DL methods are promising\cite{Kneiding15522} and yet remain largely underexplored. Semiempirical methods like g-xTB,\cite{froitzheim2025g} with accuracy approaching that of DFT, can also soon play a key role by accelerating the evaluation of the fitness. \\

\noindent
Besides GAs, one of the first attempts at using DL for TMCs was TS-GAN,\cite{Makos024116} a model that generates transition state geometries for catalytic reactions from a simple representation of its reactants and products. However, GANs, one of the earliest DL methods available, showed limitations when addressing TMC complexity and the field quickly shifted to methods learning smoother latent spaces, among them, VAEs (Figure \ref{fig:TMCs}c). The JT-VAE method\cite{Jin2018} is particularly useful for TMCs, since it facilitates the generation of complex structures via the junction tree (JT) formalism, in which molecular graphs are coarse-grained into clusters that aggregate bonds and rings. JT-VAE models can be used for the inverse design of TMCs, conditioning ligand generation by field strength.\cite{Lee1095} In order to enhance the generative process, the latent space of the model is structured via a neural network that predicts the target property; for example, the oxidative addition reaction energy, which enables the inverse design of cross-coupling catalysts.\cite{Schilter728} In these applications, SMILES\cite{weininger1988smiles} or SELFIES\cite{krenn2020selfies} strings are used either isolated,\cite{Lee1095} to represent individual ligands, or aggregated,\cite{Schilter728} to represent full TMCs. Aggregated representations can be obtained without the need to define explicit metal--ligand bonds by using the fragmentation token within the SMILES syntax. While conveniently simple, this approach leads to ambiguities if a ligand contains multiple binding moieties. Strandgaard and coworkers addressed this issue by developing a SMILES encoding of the metal anchors that supports metallacycles in TMCs that have dentic or haptic ligands.\cite{Strandgaard2294} The resulting JT-VAE enables generative tasks with high novelty, validity, and uniqueness (Box 2). These tasks can be conditioned on multiple properties, separately or jointly, including the TMC metal charge and HOMO-LUMO gap, and the ligand bulkiness and solubility. For more involved inverse designs in catalysis, the CatDRX VAE model\cite{Kengkanna2399} (Figure \ref{fig:TMCs}d) learns a latent space that encodes not only the catalyst, but also the reactants and reaction conditions, like additives, solvent, and time, adding the possibility of pre-training the model with general datasets for higher efficiency and sustainability. The capabilities of the CatDRX model have been demonstrated for both organic- and metal-catalyzed reactions, including C-C cross-coupling. Despite the significant progress made, VAEs still lack full support for whole-TMC generation, though recent work on string representations may soon solve this problem.\cite{Rasmussen63, Rusanov2026} \\

\noindent
Bypassing the challenges of representing TMCs with strings and graphs, and following the interest on generating 3D structures, which are highly valuable in applications like asymmetric catalysis, DMs are emerging as a powerful alternative to VAEs. The LigandDiff DM model,\cite{Jin4377} for example, can diversify TMCs by conditioning ligand generation in a given coordination site on all other ligands, which are fixed. This is useful in systems like catalysts and photosensitizers in which a specific ligand plays an active role, while the others are spectators. LigandDiff encodes TMCs as a point cloud of atomic numbers and positions, associating each atom to its ligand to facilitate the denoising process. After extending the original implementation for monodentate ligands,\cite{Jin4377} LigandDiff now also allows for conditioning generation on higher denticity orders, and for either a single or multiple ligands.\cite{Jin8367} In the OM-Diff DM,\cite{Cornet1793} both geometry generation and property conditioning are done at the whole-TMC level. Besides generating catalysts for cross-coupling reactions, OM-Diff can also leverage E(3)-equivariance\cite{cornet2024equivariant} to generate transition states under composition and geometry constraints,\cite{cornet20250630} as proven for H$_2$ activation in the chemical space surrounding the Vaska's complex.\cite{Friederich4584} Challenged by high computational costs, and still far from covering the many variables that define the TMCs space, DMs remain an active field of research yet in its infancy. \\

\noindent
The broad chemical knowledge encoded in LLMs can also be effectively leveraged in the inverse design of TMCs. For example, in the context of automation for autonomous chemistry labs,\cite{Tom9633} the GPT-based CoScientist model can assist the optimization of catalytic cross-coupling reactions through an iterative feedback loop.\cite{Boiko570} Tested on experimental datasets for the Suzuki–Miyaura and Buchwald–Hartwig reactions, CoScientist outperformed standard Bayesian optimization, while also providing chemical reasoning to support its recommendations. This paradigm was extended in the LLM-EO model,\cite{Lu32377} in which an LLM model generated TMCs within a GA tasked with the dual-objective optimization of their HOMO-LUMO gap and polarizability, showing the potential of hybrid DL/EC methods.\cite{Kneiding15522} \\

\noindent
The choice of a generative model for TMC inverse design should be guided by its nature, as well as by data availability. VAEs and DMs are ideal for broad exploration but require large training datasets. DMs in particular offer great flexibility in terms of ligand denticity and metal center coordination number and geometry, whereas VAEs seem to be more easy to train. While both models support conditional generation, the  amount of training data needed is usually not available or prohibitively expensive to obtain. In contrast, GAs are well suited for low-data regimes, enabling efficient multi-objective optimization. LLMs are scarce and seem best suited for exploratory studies, though their use could be generalized to serve as chemist-machine interfaces. All methods have proven their potential in complex applications like catalysis, though with limited experimental verification.

\section*{Non-porous inorganic crystals}

\bigskip

\noindent

\noindent
Besides the inverse design of molecular TMCs, generative AI has been used in the exploration of the vast chemical space spanning the non-porous inorganic crystals.\cite{Metnie23620} In addition to perovskites,\cite{Wang181496, Chenebuah198, Wu1256, Türk9455, Choubisa433} these include 2D materials,\cite{Fung200} catalytic metal surfaces,\cite{Wang147, Song1053, Xin102} nanoparticles,\cite{Rahman634, He17444, Peurifoyeaar4206} battery materials,\cite{Datta102665, Li4964} alloys,\cite{Sun54, Long66} and metal oxides.\cite{Noh1370} But beyond tailored models, there is a strong interest in models that generalize the scope of inverse design to any compact, non-porous crystalline material (Figure \ref{fig:Introduction}c).\cite{Noh4871, Wang365, Cheng174, DeBreuck370} Adding to chemical validity and Euclidean invariances, which are already critical in molecular TMCs, inorganic crystals involve one further complication: the geometric constraints intrinsic to their structural periodicity, the symmetries of the 230 possible space groups. The way and the extent to which generative AI tackles these challenges are key criteria in selecting the most appropriate method, and defines how the field has evolved from GAs to DMs, through GANs, VAES, and, more recently, LLMs (Box 1, Figure \ref{fig:timeline}). \\

\noindent
Similar to the inverse design of TMCs, GAs define a robust baseline against which modern DL models can be benchmarked, being also used to automate the generation of their training data.\cite{Noh1370, Luo254} After defining a chemical composition, the fitness function of a GA can be designed to find the structure that minimizes its energy. In this way, the GA serves the purpose of \texttt{crystal structure prediction} (CSP),\cite{Woodley2535} which was also the aim of some early DL methods.\cite{Kim1412} Fueled by the interest of the community on CSP, several research groups developed software platforms for the application of GAs to inorganic crystals, including the CALYPSO,\cite{Wang2063} USPEX,\cite{Lyakhov1172} and XtalOpt\cite{Lonie372} programs. Whereas USPEX and XtalOpt are based on a GA optimizer that can be multi-objective, CALYPSO uses a different EC algorithm (particle swarm optimization).\cite{Shami10031} Other differences include how chemical composition and structure are defined and varied over different stoichiometries and symmetries. Besides CSP, GAs and other EC algorithms have enabled the generation of crystalline compounds conditioned by multiple properties,\cite{Gao51} like the band gap of titanium oxides\cite{Chen2333} and the density of states of magnetic materials.\cite{Koshkaki124704} There is also a growing interest in hybrid DL/EC methods and applications like, for example, the exploration of VAE latent spaces with GAs in the VQCrystal model,\cite{Qiu184} for the optimization of band gaps and formation energies, and the inverse design of electro-catalysts for CO$_2$ reduction.\cite{Song1053} \\

\noindent
In GAs and other EC methods, the periodic geometry and symmetry of the crystal are part of the representation and they can be kept or modified in an exact, deterministic manner. This aligns with the first-principles computation of the fitness, usually at the DFT level, which is also the main disadvantage of this approach: high computational cost. Another issue is that these algorithms operate under composition and structure constraints that restrict the diversity of the generated compounds. These limitations motivated the development of DL methods for the inverse design of inorganic crystals, with GANs being among the earlier attempts. Notable examples include CrystalGAN,\cite{Nouira2018} MatGAN,\cite{Dan84} DCGAN,\cite{Long66} CubicGAN,\cite{Zhao2100566} and PGCGM.\cite{Zhao38} A characteristic of GANs, and all other DL methods, is the use of very large datasets; for example, MatGAN and CubicGAN were trained with over 350k crystalline materials extracted from the OQMD and MP datasets, among others. These models also differed by the quality of the representations used to capture geometry and symmetry, varying from simple 2D crystal graphs to complex tensors encoding rich crystallographic information. The latest PGCGM GAN, added symmetry constraints yielding a significant boost in the physical validity of the generated crystals.\cite{Zhao38} At the application level, CubicGAN generated novel crystal formulations conditioned by structural properties; for example, ABC$_6$D$_6$ and AB$_8$C$_{12}$ were inversely designed for space groups 216 and 221, respectively.\cite{Zhao2100566} Though DFT calculations are not needed to run GAN models, they are still used for their evaluation and for producing the training data.\\

\noindent
The competition between generator and discriminator in GANs challenges the learning of smooth latent representations of crystals. This is critical when compound generation needs to obey entangled constraints like geometry-symmetry correspondence within space group theory. VAEs address this problem via a \texttt{loss function} term that normalizes the latent space towards a Gaussian distribution. The iMatGen model\cite{Noh1370} was among the first crystal VAE attempts. Without explicit support for geometry and symmetry relations in its architecture, these properties were learned from the training data through a 3D voxel representation. Showing a performance similar to GAs, though with a higher computational cost, iMatGen was limited to the exploration of the vanadium oxide space. The Fourier transform crystal properties (FTCP) VAE\cite{Ren314} extended this scope to any ternary or quaternary material, using a rich representation including chemical composition and atomic coordinates. The low to moderate validity (Box 2) of the crystals generated by these earlier VAE models was thereafter improved by adding Euclidean invariances, and by mapping the latent space onto multiple target properties with neural networks, which also enabled complex inverse design tasks. The higher performance of this approach was proven by the Cond-CDVAE model,\cite{Xie2022, Luo254} which integrated an equivariant \texttt{graph neural network} (GNN) encoder with a DM denoiser to execute complex generative tasks resolving polymorphism. The state of the art of VAEs is defined by models that can encode and reconstruct Wyckoff positions. For example, the use of this representation in the WyCryst VAE\cite{Zhu3469} provided ample support for the crystal geometries and symmetries of all space groups, outperforming all previous VAE models. WyCryst can be used for advanced inverse design in which conditioning on band gap and formation energy is extended with a measure of synthesizability (Box 2).\cite{Zhu8210} \\

\begin{figure}[H]
    \centering
    \includegraphics[scale=0.57]{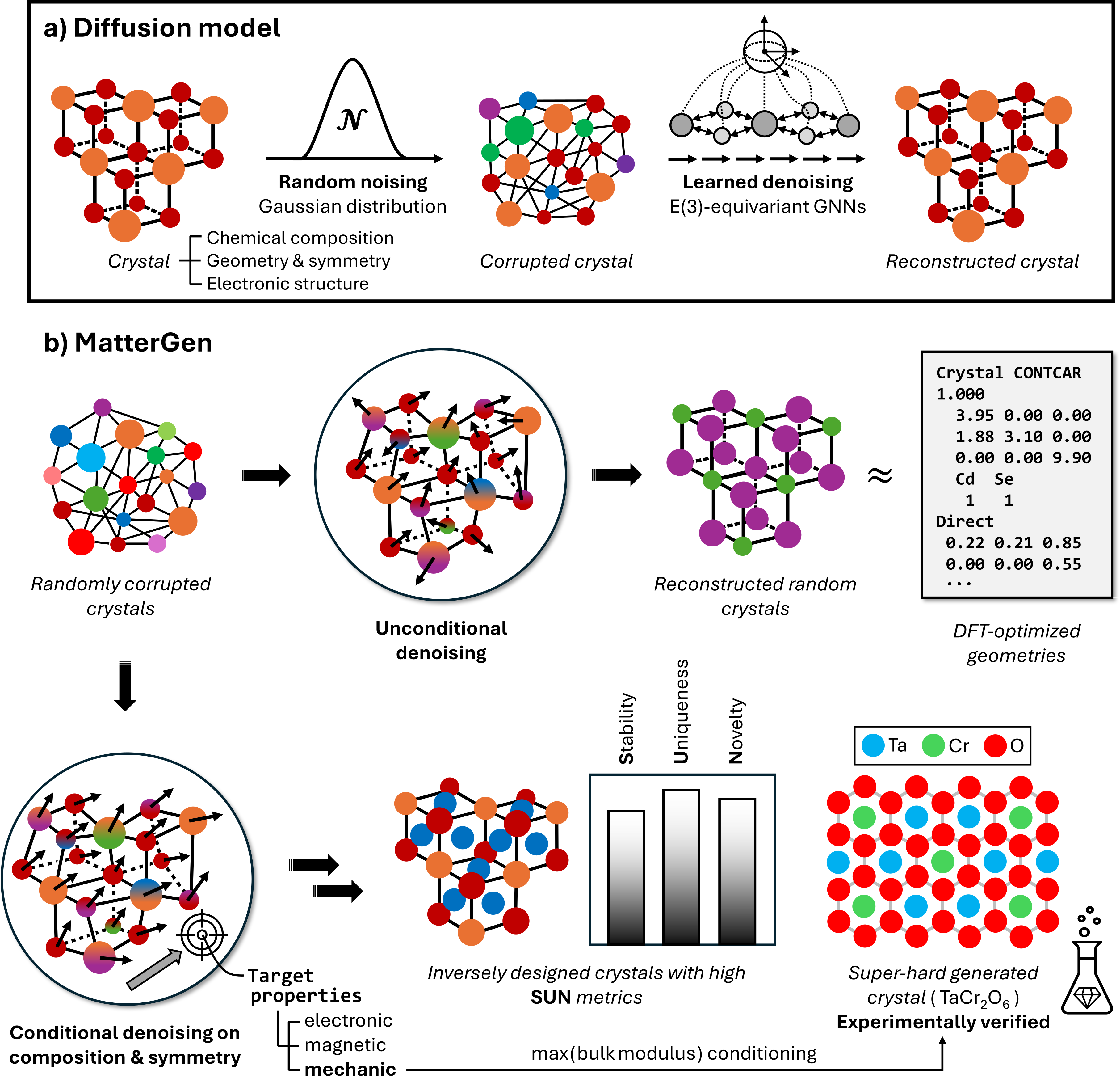}
    \caption{Diffusion models for non-porous inorganic crystals. \textbf{a)} Noising using of crystal structure representation with Gaussian noise and learned denoising with E(3)-equivariant graph neural networks; \textbf{b)} The MatterGen diffusion model doing unconditional generation with nearly DFT quality and conditional generation on multiple properties, using SUN metrics for evaluation and experiments for verification.}
    \label{fig:non-porous_crystals}
\end{figure}

\noindent
Despite their capacity of learning smoother latent spaces, VAEs struggle to define with precision the atomic positions in the crystal lattice. The generated representations can be easily reverted to 3D structures but these are often roughly approximated, requiring expensive DFT refinement. DMs tackle this issue by replacing first-principles geometry optimization by the much cheaper denoising process that is learned within the model (Figure \ref{fig:non-porous_crystals}a). The MatterGen DM\cite{Zeni624} (Figure \ref{fig:non-porous_crystals}b) handles the joint noising and denoising of chemical composition, fractional coordinates, and lattice parameters, generalizing over the periodic table and capturing the critical interdependence between these three variables to achieve top performance in stability, uniqueness, and novelty (Box 2). Similar to previous models developed for CSP,\cite{Rui2024} MatterGen uses GNNs to leverage Euclidean invariances in the denoising process. It is also an advanced model for inverse design since it can target multiple mechanical, electronic, and magnetic properties at their extremes, and with diverse chemical compositions. Covering an extensive part of the periodic table, including main group metals, the state-of-the-art performance of MatterGen was also proven in the quality of the structures generated, close to DFT, and the experimental realization of an inversely designed super-hard material, TaCr$_2$O$_6$, a rare demonstration in this field. DMs also allow to make one step further by including quantum properties in the diffusion process. ChargeDIFF\cite{Park2025} implements this approach by encoding atomic charges, making the model `electronic-aware'. A voxel representation of the charge density is encoded by GNNs in which nodes can be either atomic positions or charge densities that exchange information to learn the latent space, in an electronic structure-consistent manner. With this DM implementation, ChargeDIFF can condition crystal generation to ion migration pathways of different topology and composition, enabling the inverse design of lithium-ion battery materials.\\

\noindent
While DMs deliver top quality at the structural level, they are also slow due to the length of the stepwise denoising, which can involve hundreds of iterations. Flow models, like CrystalFlow,\cite{Luo9267} can be regarded as a type of DM that accelerates this process by learning a differential equation that does the denoising. LLMs can also be significantly faster than DMs and thus an attractive alternative. Further, LLMs can be conveniently prompted for the inverse design of chemically and structurally complex inorganic crystals using natural language. In a co-intelligent framework,\cite{Balcells16412} any chemist can start a conversation giving feedback and context to a model that proposes materials satisfying user-defined target properties. A key mechanism in this approach is the derivation of a language-compatible representation of CIF information. Besides containing all information that precisely defines a crystal, CIF files can be easily sequenced into strings. However, the way in which these strings are broken into fragments (\emph{tokens}) is not obvious and yet it has a strong impact on how efficiently and correctly LLMs learn the underlying chemical grammar. The CrystaLLM\cite{Antunes10570} showed that the digit-by-digit tokenization of the CIF, instead of breaking it into irregular fragments, allows the model to learn efficiently the chemistry principles underlying the crystal structures. CrystaLLM was built around a small 200M-parameter core, which is more convenient than using massive frontier LLMs like the 70B-parameter LLaMA-2 model.\cite{Gruver2024} In line with the VAE and DM models, LLM performance increases with geometry and symmetry awareness. Following this direction, CrystalFormer\cite{Cao3522} was trained with crystallographic information expressed with Wyckoff positions to generalize the conditional generation of inorganic crystals, increasing both their diversity and stability. Adding text prompting to a DM, the Chemeleon model\cite{Park4379} blurred the boundary between these models and LLMs, aligning textual and structural information to enable the generation of advanced materials, like quaternary Li$_x$P$_y$S$_z$Cl$_w$ solid electrolytes for batteries. For more complex multi-property conditioning, GAs can be integrated into LLMs.\cite{Gan2025} \\

\noindent
Following the evolution of generative AI for non-porous inorganic crystals, GAs may still be the best choice for multi-property designs, if computing resources for intensive DFT calculations are available. This and other EC methods also have the advantage of relying on various software tools with a long, established background. With a much lower computational cost, once trained, VAEs can be well suited if there is no need for precise atomic coordinates, naturally providing more diverse generation than GAs. Finally, DMs define the state-of-the-art in terms of structure precision, and their recent integration into LLMs give the advantage of user-machine interactions through natural language.

\section*{Microporous inorganic materials}

\bigskip

\noindent
The use of generative AI methods for non-porous crystals is generally limited to small unit cells, in the order of $10^1$ atoms. For porous materials, this order of magnitude increases to $10^2$-$10^3$, which may require adapted strategies like coarse-grained representations in which atoms are aggregated. Further, the modeling of pore confinement effects requires the integration of additional features within the models. This has motivated the development of methods tailored for the inverse design of MOFs and zeolites (Figure \ref{fig:Introduction}b).\cite{Xie2336, Jablonka161, Park3727, Jablonka8066, Wu020601, Gandhi100739, Ozcan23367, Moghadam121} Similar to the progression made for non-porous materials, the field has evolved from early GAN models to DMs supporting Euclidean invariances (Box 1, Figure \ref{fig:timeline}). More recently, LLMs have leveraged natural language processing for conditional generation or as a user interface for other methods, like GAs, which remain a strong benchmarking reference. \\

\noindent
At the structural level, TMCs and MOFs have a similar modular nature when considering these analogies: metal center $\leftrightarrow$ node, and organic ligand $\leftrightarrow$ linker (Figures \ref{fig:Introduction}a and \ref{fig:Introduction}b). Thus, MOFs are also naturally expressed as chromosomes, enabling the use of GAs in their inverse design. GA applications have focused on the optimization of MOFs maximizing gas adsorption in green chemistry applications involving storage and separation processes. Chung and coworkers, for example, developed a GA to find top-performing MOFs in the capture of pre-combustion CO$_2$.\cite{Chunge1600909} With a simple six-digit chromosome representation, encoding the metal node, linker, and functional groups, this model allows for dual-objective Pareto optimizations balancing CO$_2$ working capacity with CO$_2/$H$_2$ adsorption selectivity. Using \texttt{grand canonical Monte Carlo (GCMC)} simulations of gas adsorption to compute the fitness, the GA only needed to evolve ten generations to identify the $4\%$ elite MOFs. After filtering for synthetic feasibility, the experimental characterization and testing of NOTT-101/OEt, the top MOF optimized by the GA, outperformed previously known materials like Mg-MOF-74. Besides gas capture, similar approaches have focused on the joint optimization of selectivity and permeability for the separation of gas mixtures, like ethylene+ethane, using GAs that explore massive MOF spaces encompassing quadrillions of metal-linker-topology combinations.\cite{Zhou256} \\

\noindent
Besides whole-MOF optimization, GAs can be used on the inverse design of the framework components. For example, MOFF-GA\cite{Collins1600954} is a chemical functionalization method that evolves organic linkers for increased CO$_2$ adsorption capacity, using a fitness function that also accounts for surface area and energy consumption. With a chromosome encoding the functional groups installed in the linker and similarity-aware genetic operations, MOFF-GA can augment the CO$_2$ capture capacity of known MOFs, like MIL-47, by a factor of four. A less conventional and yet efficient use of GAs consists in evolving multiple generations in parallel, starting from different initial populations across various topologies. Boosting MOF diversity, this approach enables multi-objective tasks in which CO$_2$ working capacity is optimized jointly with selective adsorption relative to N$_2$.\cite{Pham4585} \\

\noindent
The conditional generation of MOFs is one the fields that has proven the potential of synergizing DL and GA methods.\cite{Kneiding15522} Lee and coworkers used GCMC calculations to train MOF-NET, a neural network that predicts the work capacity fitness needed to evolve MOFs for methane storage.\cite{Lee23647} A key tool at the core of this approach is PORMAKE, an automated assembler of MOF structures in diverse topologies, based on an extensive library of metal node and organic linker building blocks. PORMAKE is conveniently available as a Python library applicable to any inverse design project focused on porous frameworks. GAs can also self-evolve by using mutation and crossover operations that adapt on-the-fly, responding to the evolution process. The Tangent Adaptive Genetic Algorithm (TAGA) implements this approach and uses ML methods based on decision trees to accelerate the computation of fitness functions accounting for selective gas uptake from mixtures like N$_2$ + methane.\cite{Lie13146} \\

\noindent
The complexity of GAN models and, particularly, the large training data they require and the difficulty of training them, has favored applications to zeolites over MOFs, owing to their lesser complexity (Figure \ref{fig:Introduction}b). ZeoGAN\cite{Kimeaax9324} uses a 3D-grid representation of both geometry and energy. Whereas the first encodes the spatial coordinates of the aluminosilicate building units, the second accounts for the potential energy of methane within the structure, which is derived from molecular dynamics (MD) simulations. In  unconditional generation, only 8 out of $10^6$ generated zeolites were confirmed to be valid, unique, and novel (Box 2), reflecting the many constraints that define this chemical space. ZeoGAN can also be used for inverse design, conditioning zeolite generation to a target range of methane adsorption energies. Making a logical step towards the generation of 3D geometries, Park and coworkers developed the ZeoDiff DM model.\cite{Park6507} Using similar representations and training data, ZeoDiff outperformed ZeoGAN by three orders of magnitude in terms of structural validity, also extending inverse design tasks with conditioning on void fraction and Henry constant. \\

\noindent
In MOF design, DM models were preceded by VAEs. The supramolecular VAE (SmVAE) model\cite{Yao76} enabled the conditional generation of MOFs with improved selectivity in CO$_2$ adsorption from mixtures including other gases like N$_2$ or methane. Trained on the CORE MOF database,\cite{Chung5985, Zhao102140} this model uses a reversible representation encoding the topology of the framework together with the identity of the metal nodes and organic linkers. Beyond VAEs, and considering the complexity of MOFs, a sensible entry to DMs is to focus the noising-denoising process on only one structure component. DiffLinker\cite{Igashov417} can do this for the organic linkers, though it was originally developed for drug discovery conditioned on protein pocket shapes. Proving the potential of cross-fertilization in this field, the GHP-MOFassemble method\cite{Park21} leverages GCMC simulations for the HTVS of large spaces combining DiffLinker-generated linkers with metal nodes from the hMOF dataset,\cite{Wilmer83} finding novel MOFs with high CO$_2$ adsorption capacities. \\

\newpage

\noindent
Further research on DMs for MOFs yielded the MOFDiff\cite{fu2023mofdiff} and MOFFUSION\cite{Park34} models. MOFDiff leverages a 3D coarse-grained representation of the structure that reflects both the modular and topological structure of the MOFs. Trained with a large 300k dataset,\cite{Boyd253} this DM can be used for conditioning MOF generation on CO$_2$ capture capacity. The atomic structure of the frameworks is recovered by an assembly algorithm that matches connecting points between nodes and linkers, followed by geometry optimization with a force field. A final validation check done with the MOFChecker code\cite{Xin1560} yielded a success rate of 30\%. More recently, MOFFUSION advanced the state-of-the-art of DM models by integrating a builder module that predicts structural components from generated MOF representations, which are then assembled into frameworks using PORMAKE.\cite{Lee23647} A distinct feature of this DM is the use of a signed distance function that yields a continuous representation of the microporous structure adapted to the denoising process. With an 81\% validity rate, MOFFUSION outperformed the previous DM models, while enabling inverse designs aimed at multiple targets, including pore volume, framework topology, and metal node, accepting also text prompts. \\

\noindent
By revalorizing and putting data at the center, GANs, VAEs, and DMs set the stage for the emergence of LLMs. Adding to this, and in line with the exponential growth of research in the field, there has been intense research on text-mining the extensive MOF bibliography to extract relevant information, like surface area, pore volume, synthesis protocols, and thermal stability, while assessing its value in the training of discriminative AI models.\cite{Park244, Bae11083, Luoe202200242, Nandy17535, Nandy74} The next natural step was to use recurrent neural networks to encode all this text data and to leverage it in conditional generation tasks.\cite{Jensen858, Xu119188, Cleeton4806} With the advent of the transformer architecture, which enabled the training and deployment of LLMs at scale, Zheng and coworkers made the next natural step by developing a GPT-based LLM for consulting synthesis recipes and doing data analysis.\cite{Zheng369} Through a co-intelligent approach,\cite{Balcells16412} this model was guided with ChemPrompt, a prompt engineering concept based on chemistry principles that minimizes hallucination while providing detailed instructions and requesting structured replies.\cite{Zheng18048} \\

\begin{figure}[H]
    \centering
    \includegraphics[scale=1.85]{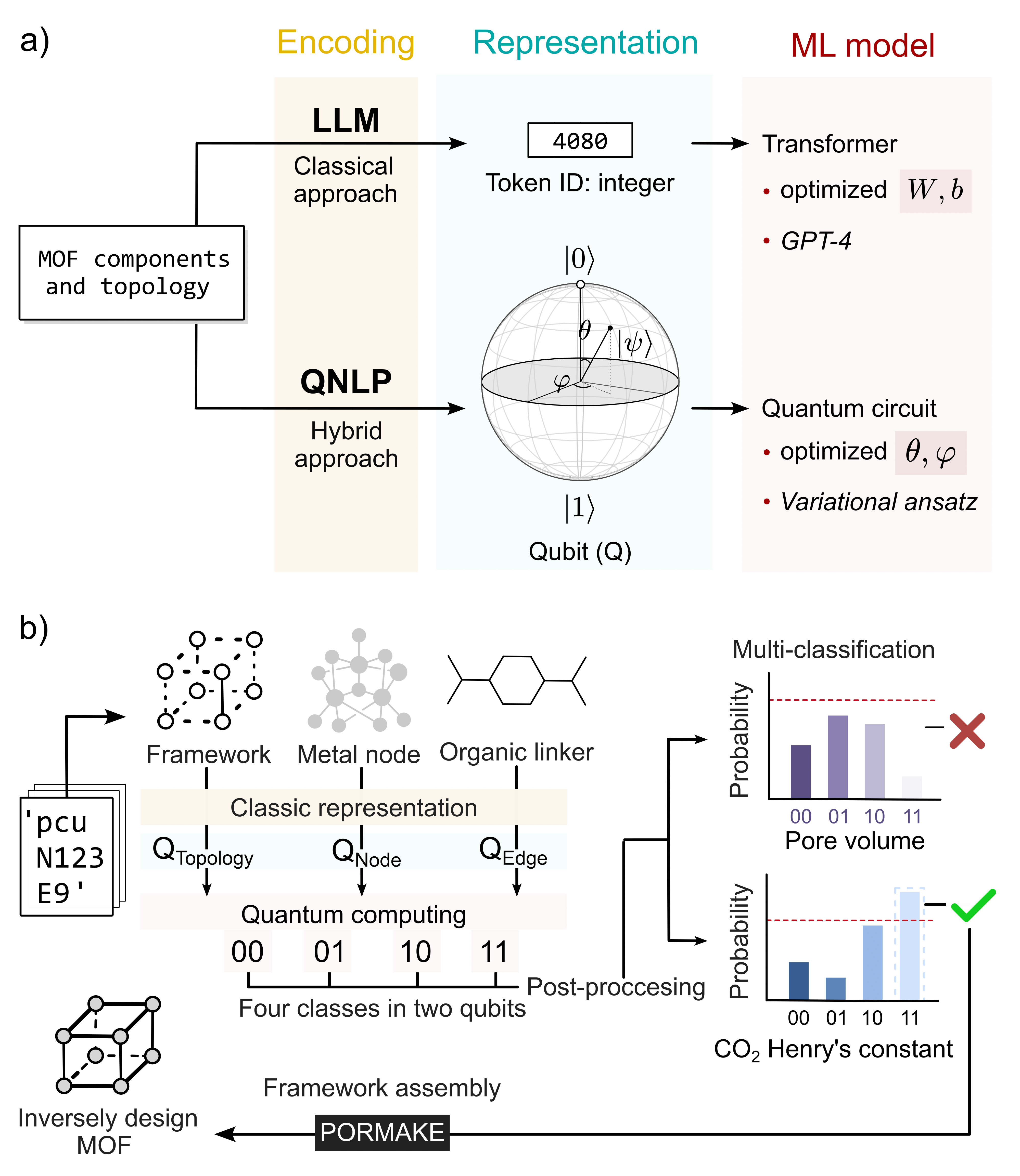}
    \caption{LLM models for MOFs. \textbf{a)} Whereas the classical approach encodes the materials information with tokens, quantum natural language processing (QNLP) uses a qubit representation. \textbf{b)} The generative process leverages a four-category classification task (low, moderate low, moderate high, and high) over two target properties (pore volume and CO$_2$ Henry's constant).}
    \label{fig:microporous}
\end{figure}

\noindent
After these precedents, the ChatMOF LLM\cite{Kang4705} pushed the field further by integrating automated data retrieval, property prediction, and inverse design with large frontier models like GPT-4. The model is built around three components that interact with the user in a closed loop: an agent that defines an action plan from the user prompt, a toolkit that executes the research plan, and an evaluator that assesses and gives the output. MOF inverse design was implemented with natural language-operated GAs that assemble the frameworks with PORMAKE\cite{Lee23647} and compute the fitness via GCMC simulations. These GAs can condition generation on single targets, like surface area, or multiple targets, like H$_2$ adsorption capacity at a given pressure and temperature. Besides inverse design, ChatMOF can also be used for search and prediction, with these three tasks achieving accuracies of 70, 93 and 90\%, respectively. The future of LLMs for MOFs and materials in general may be strongly rooted in quantum technologies, which have the potential of bringing massive gains in speed and the size of the chemical spaces that can be explored.\cite{Kang1921} Kang and Kim have recently shown how quantum natural language processing (QNLP) can be integrated into an LLM for MOF inverse design (Figure \ref{fig:microporous}a).\cite{Kang321} For each of the two target properties, which were the pore volume and the CO$_2$ Henry's constant, four different classes are encoded as $(00, 01, 10, 11)$ using two qubits (Figure \ref{fig:microporous}b). Based on this representation, the QNLP model leverages a hybrid architecture, with a classic part computing the loss function and assembling the frameworks, and a quantum part learning the correlation between the four classes and the parameters defining the MOFs (topology, node, and linker), achieving high accuracy in conditional generation tasks. \\

\noindent
LLMs can also be used to guide the selection of appropriate organic structure-directing agents (OSDAs) for the synthesis of zeolites. The most common OSDAs are amines that facilitate zeolite formation while orienting their assembly towards a specific topology. In contrast with the zeolite chemical space, which is constrained to a much smaller size than that of the MOFs, the OSDA space has a size similar to that of the drug-like molecules, $\sim10^{60}$, requiring AI methods for HTVS exploration.\cite{Balcells608, Xie661} Ito and coworkers developed an LLM platform based on the GPT-4 model that suggests OSDAs in SMILES notation, learning from empirical information and atomistic simulations data.\cite{Ito2447} The LLM output is filtered using parameters like structural rigidity, chemical stability, and the ratio between carbon and nitrogen atoms. Failed designs can be passed to the model as feedback to improve its performance. For inverse design tasks, this LLM has proven its capacity in the generation of stable OSDAs for various zeolite types. \\

\noindent
An important criterion for selecting the appropriate AI method is whether the design needs to be holistic or component-centered. GAs can conveniently do both, for either MOFs or zeolites, offering other advantages, like conditioning generation on multiple targets. However, they require expensive GCMC simulations for the computation of the fitness, though DL-accelerated surrogate models are already available. For zeolites, early GAN models have been clearly outperformed by DMs, which are also flexible in the whole- versus component-focus of the design. Similarly, for MOFs, DMs outperformed earlier VAE models, and the state-of-the-art already enables multi-target designs. Finally, LLMs can act as chemical assistants that facilitate design tasks via the consultation of relevant information like synthesis recipes. Supporting both MOFs and zeolites, LLMs may also have a readiness advantage in the use of upcoming technologies related to quantum information theory.

\newpage

\newpage

\section*{Outlook}

\bigskip

\noindent
Despite significant achievements, the field of generative AI for inorganic compounds is yet in its infancy, far from realizing the inverse design of complex systems, like polynuclear TMCs and crystal defects, or challenging environments, like excited states and non-equilibrium conditions. Aiming at this goal, the community has to develop new methods yielding higher quality results with higher computational efficiency. \\

\noindent
One factor hampering the further development of the field is the lack of generally accepted benchmarks. Recent studies have proposed different directions towards this end,\cite{Szymanski8000, Yong1889, Betala2025} but remain far from being as established as those set for organic molecules, which have a longer tradition.\cite{brown2019guacamol, nigam2023tartarus, Polykovskiy565644} A key element in this endeavour is the adoption of a comprehensive evaluation system that considers all metrics collected in Box 2, some of which are neither fully developed yet, nor transferable over the wide diversity of the inorganic compound space. \\

\noindent
The \texttt{SUN metric} was recently defined for the evaluation of the MatterGen DM model,\cite{Zeni624} putting the focus on stability, uniqueness, and novelty (SUN). Other similar but not fully consistent metrics have been reported, setting a framework in which it is difficult to compare the models for their further improvement. SUN is a robust evaluation metric but it can be insufficient in several scenarios. Stability, for example, is important but often ill-defined for crystals, since common criteria, like the energy above the \texttt{convex hull}, cannot define a universal limit between stability and metastability. DFT phonon analysis for the detection of imaginary modes would be more robust,\cite{Han2025} but it is computationally expensive. Besides, and regardless of (meta)stability, chemical and physical validity can be a better, complementary metric, but its implementation is much easier for molecules than for periodic materials, with recent efforts focused on MOFs\cite{Xin1560} showing the detrimental effect that this can  have on the quality of the training data.\cite{White17579} Uniqueness is also critical since it quantifies the generative efficiency of the model. It should be carefully defined in relation to novelty to avoid issues like the presence of small, irrelevant variations of systems that are either generated or already present in the training data.\cite{Cheetham3490} Similarity measures are thus important from this perspective, playing also an important role in quantifying diversity. Novelty is fundamentally relevant to the value of the model and it can be re-formulated from the perspective of rediscovery, often ignored despite its value in assessing transferability to neighbouring chemical spaces. \\

\noindent
Beyond the SUN metric, data augmentation (Box 2) is rarely reported despite its relevance in unconditional generative tasks aimed at the creation of virtual libraries or synthetic data for training DL models after labelling.\cite{Chen128167, Li2025} In this regard, diversity plays a key role in assessing the quality of the data, particularly in EC algorithms, given their tendency to get trapped into local optima. In conditional generation for inverse design, verification is the critical measure of success. Verification can be done by computing the target property at the same level of theory used for the training data, often DFT. However, from a more practical standpoint, it is the experimental verification of the inverse designs what is most important and, yet, examples of such verification remain rare.\cite{Zeni624}

\begin{tcolorbox}[width=\textwidth, colback=white, title={\textbf{Box 2. List of evaluation metrics for a systematic assessment of generative AI methods.}}, colbacktitle=Maroon, colframe=Maroon, coltitle=white, arc=2mm, boxrule=0.75mm, boxsep=1.75mm]

\bigskip

\textbf{\underline{SUN metrics}:}

\noindent
\textbf{Stability.} The ratio of generated compounds that are stable or metastable relative to a quantitative measure like a DFT energy threshold above the convex hull.

\noindent
\textbf{Uniqueness.} The ratio of generated compounds that do not appear repeated within the output of the model.

\noindent
\textbf{Novelty.} The ratio of generated compounds that are novel relative to those used to train the model.

\bigskip

\textbf{\underline{Beyond SUN metrics}:}

\noindent
\textbf{Rediscovery.} The number of generated compounds that are equal to other known compounds outside the training dataset and thus not seen by the model.

\noindent
\textbf{Validity.} The ratio of generated compounds that meet the physical principles underlying their chemical composition and structure. A valid compound can be either stable or metastable.

\noindent
\textbf{Synthetic data augmentation.} The factor measuring the extent to which the size of the data can be increased for data synthesis applications.

\noindent
\textbf{Diversity.} Averaged measure of similarity among the generated compounds based on a quantitative index like the Tanimoto coefficient or the Cosine distance.

\noindent
\textbf{Verification.} In inverse design, the ratio of generated compounds that fulfil the target property.

\noindent
\textbf{Synthesizability.} Averaged measure of synthetic accessibility among the compounds generated by the model.

\end{tcolorbox}

\newpage

\noindent
The main barrier between computational generation and experimental verification is synthesizability, which is poorly correlated with stability. In principle, this critical variable can be simply addressed as one more target conditioning the generation. However, whereas synthesizability for organics can be defined as a continuous measure on the basis of database statistics and measures of similarity and molecular complexity,\cite{Ertl8, Gao5714} inorganics are more challenging. For example, for TMCs, synthesizability is easily determined for the organic ligands but not for their assembly around the metal center. Besides early, naive approaches, like synthesizable/not synthesizable classification corresponding to absence/presence in datasets like the ICSD,\cite{Ren314} recent work on inorganic materials has focused on binary ML classifiers,\cite{Antoniuk155, Jang18836, Zhu8210} considering the implicit biases in the training data.\cite{Davariashtiyani040501} Building on chemoinformatics tools for organic compounds could be an efficient strategy, specially for TMCs, for which the adaption of string representations like SMILES\cite{Rasmussen63} and TUCAN\cite{Brammer66} can be particularly useful. However, a universal synthesizability measure for inorganics should go beyond data statistics and complexity heuristics, being also retro-synthesis aware\cite{Guo6943} and leveraging LLMs\cite{Song6530, Choi39113} in chemist-machine interfaces. \\

\noindent
Cross-fertilization over different inorganic compound types (Figure \ref{fig:Introduction}) and AI methods (Box 1) should also provide transformative opportunities to the inverse design of inorganic compounds. Predictive AI methods like MLIPs,\cite{Kalita1120} can accelerate the computation of rigorous measures of stability (Box 2), like phonon analysis,\cite{Loew178} and their formulation for materials like MOFs\cite{Krab4} can facilitate applications to other systems, including TMCs. MLIPs may also open fruitful venues in the generation of non-equilibrium structures,\cite{Park71046} and the combination of methods like MatterGen\cite{Zeni624} and MatterSim\cite{Yang2024} can integrate the exploration of the chemical and configurational spaces in TMCs and MOFs, when combinatorial explosions involve flexible fragments. For the synergy between EC and DL methods, GAs can already augment LLMs by adding multi-objective optimization tasks\cite{Lu32377} and explore latent spaces,\cite{Qiu184} while neural networks accelerate the computation of the fitness function through DFT surrogates.\cite{Kneiding15522} Integrating DL with EC to achieve both optimality and novelty in inverse design has been shown for TMCs\cite{pita2025evolving} and may also have potential in crystalline materials. For crystal defects and other unusual but highly relevant structural features, local cluster models can leverage the generative AI methods developed for TMCs. Regarding lab automation,\cite{Oh157, Tom9633} GAs may open new opportunities for high-throughput experimentation (HTE) that can be broadly relevant for any type of inorganic compound. The similar size of GA populations and HTE batches should allow for complementing computational evolution with experimental verification. \\

\noindent
Pushing forward the state-of-the-art of the field, next-generation models should expand inverse design towards exceptional and out-of-distribution compounds.\cite{Schrier21699, Omee144} Chemical space exploration should be extended to non-equilibrium structures and their relation to crystal defects and disorder,\cite{Jakobe14226} amorphous materials,\cite{Liu228} and the excited states of both molecular and periodic systems.\cite{Westermayr9873, Kneiding11766, Ren314} In TMC space, the generation of polynuclear complexes conditioned on the intricate relations between charge and spin in multi-metallic centers is also uncharted territory. In crystal space, more efficient methods would allow expanding the unit cell size to the point at which the non-porous models, with their rich composition-symmetry-electronic encodings, become available for the inverse design of porous materials. \\

\noindent
Higher efficiency will also translate into higher sustainability,\cite{Sandonas2026} a widespread issue in generative AI, given the extensive use of water and power in the supercomputing centers used to generate and process the training data, and to optimize and deploy the models. One approach that has a direct impact on sustainability is the pre-training of an expensive general model followed by its fine-tuning, which, at a much lower computational cost, yields multiple models solving specific tasks. This concept has proven its potential for organic molecules\cite{yu2021review, Blaschke363, Nafiz154} but remains seldom explored for inorganic compounds.\cite{Badrinarayanan9049} Also in the context of sustainability, the further development of the field has the potential of accelerating the discovery of new catalysts and energy materials reducing waste generation and energy consumption. \\

\noindent
Combined with improved benchmarking, these method developments and their integration will improve the quality and efficiency of generative AI towards the inverse design of inorganic compounds of higher complexity, extending actual applications to catalysis and energy materials, and creating new opportunities in emerging fields like metallodrug discovery.

\section*{Acknowledgments}
D.B. acknowledges the RCN FRIPRO program for ground-breaking research (catLEGOS project; number 325003). L.M.-G. thanks the EU Commission for support through the HE-MSCA-PF program (BiFUCCO2 grant; number 101206288). N.K. acknowledges the funding from the European Research Council (ERC) under the EU Horizon 2020 program (CuBE grant; number 856446). All authors acknowledge the RCN program for the National Centers of Excellence (Hylleraas Centre; project number 262695).

\section*{Author contributions}
H.K., L.M.G., N.K., A.N., and D.B. contributed to scientific discussions and to the reviewing and editing of the manuscript. H.K., L.M.G., N.K., and D.B. contributed to researching data for the manuscript and its writing. D.B. designed the Perspective concept. 

\section*{Competing interests}
The authors declare no competing interests.

\end{document}